\documentclass[journal,draftclsnofoot,onecolumn,12pt]{IEEEtran}
\usepackage{cite}
\usepackage{amsmath,amssymb,amsfonts}
\usepackage{algorithmic}
\usepackage{graphicx,color}
\usepackage{textcomp}
\usepackage{float}
\usepackage{array}
\newcolumntype{P}[1]{>{\centering\arraybackslash}p{#1}}
\newcolumntype{M}[1]{>{\centering\arraybackslash}m{#1}}
\usepackage{amsthm,amssymb,amsmath,graphicx,multirow,color,amsfonts}%

\usepackage{cite}
\usepackage[justification=centering]{caption}
\usepackage{textcomp}
\usepackage{psfrag}
\usepackage{arydshln}
\usepackage{url}
\usepackage{soul}
\usepackage{graphicx,color}
\usepackage[nolist]{acronym}
\usepackage{algorithm,algorithmic} 

\usepackage{mathtools,lipsum}
\usepackage{cuted}
\usepackage{amsmath}
 
\usepackage{mathrsfs}

\usepackage[capitalise]{cleveref}
\Crefname{equation}{Eq.\!}{Eqs.\!}
\Crefname{figure}{Fig.\!}{Figs.\!}
\Crefname{tabular}{Tab.\!}{Tabs.\!}
\Crefname{section}{Section\!}{Sections.\!}

\begin{document}


\title{Exploring UAV Networking from the Terrain Information Completeness Perspective: A Tutorial}

\author{
Zhengying~Lou, Ruibo~Wang, \\ Baha~Eddine~Youcef~Belmekki,~\IEEEmembership{Member,~IEEE}, \\ Mustafa~A.~Kishk, {\em Member, IEEE}, and Mohamed-Slim~Alouini, {\em Fellow, IEEE}
\thanks{
Zhengying Lou, Ruibo Wang, Baha Eddine Youcef Belmekki and Mohamed-Slim Alouini are with King Abdullah University of Science and Technology (KAUST), CEMSE division, Thuwal 23955-6900, Saudi Arabia. Mustafa A. Kishk is with the Department of Electronic Engineering, Maynooth University, Maynooth, W23 F2H6, Ireland. (e-mail: zhengying.lou@kaust.edu.sa; ruibo.wang@kaust.edu.sa; bahaeddine.belmekki@kaust.edu.sa; mustafa.kishk @mu.ie; slim.alouini@kaust.edu.sa).}
\vspace{-1cm}
}


\vspace{-8mm}

\maketitle

\begin{abstract}
Terrain information is a crucial factor affecting the performance of
unmanned aerial vehicle (UAV) networks. As a tutorial, this article provides a unique perspective on the completeness of terrain information, summarizing and enhancing the research on terrain-based UAV deployment. In the presence of complete terrain information, two highly discussed topics are UAV-aided map construction and dynamic trajectory design based on maps. We propose a case study illustrating the mutually reinforcing relationship between them. When terrain information is incomplete, and only terrain-related feature parameters are available, we discuss how existing models map terrain features to blockage probabilities. By introducing the application of this model with stochastic geometry, a case study is proposed to analyze the accuracy of the model. When no terrain information is available, UAVs gather terrain information during the real-time networking process and determine the next position by collected information. This real-time search method is currently limited to relay communication. In the case study, we extend it to a multi-user scenario and summarize three trade-offs of the method. Finally, we conduct a qualitative analysis
to assess the impact of three factors that have been overlooked in terrain-based UAV deployment.
\end{abstract}



\maketitle

\section{Introduction}\label{section1}
\subsection{Motivation}
In urban areas, factors such as high population density, towering buildings, and complex terrain may pose challenges to the stability and coverage of wireless communication networks \cite{ansari2021urban,zaid2023aerial}. Moreover, in places with large gatherings and high population mobility, such as sports stadiums and event venues, ensuring the reliability and stability of network connections is also an important issue \cite{wang2023terrain,huang2023system,wang2024ultra}. It is foreseeable that future networks will achieve more granular and seamless coverage \cite{imran2019seamless}. Furthermore, with the real-time changes in user demands, network services should be adaptive and self-evolving \cite{letaief2019roadmap}. The UAV network is one of the best solutions to meet these advanced communication requirements \cite{fan2022learning,shang2019unmanned}. It can assist ground networks in providing coverage for wireless network blind spots. Additionally, it is expected to serve forthcoming aerial users and vehicles, including electric vertical take-off and landing (eVTOL) aircraft, in urban air mobility (UAM) \cite{zaid2023evtol}.  At the same time, the mobility of UAVs enables the network to be deployed in a real-time and flexible manner according to actual needs \cite{fan2020robust}.

\par
However, UAV communication also faces significant limitations. Since UAVs are often used for temporary and dynamic coverage, they frequently cannot obtain a stable and powerful power supply \cite{shang2018wireless}. Therefore, UAVs are unable to sustain high signal transmission power, and as a result, they often operate at lower flight altitudes to provide better network services by approaching users closely. Given their relatively low flight altitudes, UAVs are often deployed in the gaps of obstacles' shadowing, casting frequent blockage of line-of-sight (LoS) UAV-user link \cite{gomez2013performance}. Unfortunately, due to the limited transmission power of UAVs, if the LoS is obstructed by blockages such as trees, the received signal at the user is unstable and experiences significant attenuation. The above reasons make terrain one of the most critical factors affecting UAV deployment \cite{khuwaja2018survey}.

\begin{table*}[]
\centering
\caption{Summary of terrain-based UAV networking studies}
\label{table1}
 \resizebox{\linewidth}{!}{
 \renewcommand{\arraystretch}{1.2}
\begin{tabular}{|c|c|c|c|c|c|}
\hline
Ref. & Completeness & Topology & Stage & Research objective & Application   scenario \\ \hline
\cite{esrafilian2018learning} & Complete   & Single-hop & Hybrid  & Optimize channel parameters learning  &  UAV-aided wireless network \\ \hline
\cite{ huang2021navigating} & Complete   & One-to-many & Hybrid  & Minimize the   distances between UAVs and targets & Monitor moving targets \\ \hline
\cite{ mou2021deep }        & Complete   & One-to-many & Hybrid  & Optimize coverage and reduce overload                  & Cover irregular terrain surface  \\ \hline
\cite{yi2022joint} & Complete   & One-to-many & Hybrid  & Improve the coverage and capacity &  Establish links from a BS to users \\ \hline
\cite{esrafilian2020three} & Complete   & One-to-many & Hybrid  &  Optimize the performance of localization  &  Localizing outdoor ground users \\ \hline
\cite{al2020probability} & Incomplete   & Single-hop & Offline  & Predict the geometric LoS probability                  &  Different urban environments \\ \hline
\cite{armeniakos2020sir} & Incomplete   & Single-hop & Offline  & Evaluate the performance of UAV network                  & Finite interference nodes  \\ \hline
\cite{chen2023correlation} & Incomplete   & Relay & Offline  &   Enhance system performance analysis    &   Hybrid cellular networks  \\ \hline
\cite{yin2023air} & Incomplete   & Relay & Offline  &  Optimize the placement of relay UAVs   &   UAV swarm application \\ \hline
\cite{wang2023resident} & Incomplete   & One-to-many & Offline  &  Maximize coverage and energy efficiency                &  Centralized town \\ \hline
\cite{al2014optimal} & Incomplete   & One-to-many & Offline  & Maximize wireless coverage on the ground                 &  Rapid deployable relief networks \\ \hline
\cite{alzenad20173} & Incomplete   & One-to-many & Offline  & Maximize coverage and reduce consumption                 &   Assist the terrestrial network \\ \hline
\cite{ he2018towards }      & Incomplete & One-to-many & Offline & Optimize   coverage and connectivity  & Provide emergency communication    \\ \hline
\cite{zhang2020radio}      & No information & Single-hop & Real-time & Maintain communication with base station &  Complete flight mission    \\ \hline
\cite{dong2022radio}      & No information & Single-hop & Real-time & Ensure link quality on the shortest path    & Anti-jamming communication   \\ \hline
\cite{chen2019efficient}      & No information & Relay & Real-time & Boost capacity with linear search length  &  Provide coverage extension  \\ \hline
\cite{zheng2022geography}      & No information & Relay & Real-time & Develop search strategies for UAV placement   &  Densely obstructed areas  \\ \hline
\cite{zheng2023online}      & No information & Relay & Real-time & Optimize the performance of relay links &  Construct relay links to two users \\ \hline
\end{tabular}}
\end{table*}

\subsection{Challenges of Terrain Information Utilization}
Except for a few scenarios in remote areas with expansive visibility, a large amount of literature on UAV communication networks has taken the influence of terrain into account. In summary, terrain-based UAV deployment is a complex problem that primarily involves the following three challenges. 

\par
Firstly, due to the highly complex topology of buildings, the UAV deployment should be fine-grained. Once the UAV or user has a few meters of movement, they may disappear within each other's LoS region \cite{gesbert2022uav}. Secondly, as UAVs approach users by reducing their altitude, the distance to obstacles also decreases accordingly \cite{al2014optimal}. In other words, UAVs often incur a higher probability of being blocked in exchange for reduced path loss. Lastly, when UAVs prioritize providing better service to a few selected users by altering deployment, it can potentially result in a degradation of communication performance for other users. { In summary, relevant studies focus on scenarios with different levels of terrain information completeness and address different challenges. 

\par
There are several surveys and tutorials delving into a wide range of topics related to UAV communication, including channel modeling \cite{khuwaja2018survey}, integration with 5G and future networks \cite{li2018uav, zeng2019accessing, geraci2022will}, and enhancements through artificial intelligence \cite{hashesh2022ai}. However, none of these studies comprehensively explore UAV deployment issues from the perspective of terrain information completeness. Due to the widespread interest in terrain-based UAV deployment, it is necessary to provide a comprehensive tutorial of the diverse research conducted on this topic. 
}

\subsection{Contribution}
{  From the unique perspective of terrain information completeness, this article provides a comprehensive tutorial and bibliographic index for readers interested in the field of terrain-based UAV deployment. The core contents and contributions of the article are outlined as follows:
\begin{itemize}
\item In Sec.~\ref{section2}, we categorize the existing studies into three classes from the perspective of prior terrain information completeness.
\item In Sec.~\ref{section3}, we review two pivotal aspects of the UAV-aided terrain construction method and the UAV-based trajectory design with complete terrain information. Then, a case study, for the first time, integrates these two approaches, drawing meaningful conclusions based on the interaction between these methods. 
\item In Sec.~\ref{section4}, we review the air-to-ground LoS probability model (A2GLPM), summarize how existing studies utilize this model in UAV deployment with incomplete terrain information. Then, a case study is proposed to validate the accuracy of the model. 
\item In Sec.~\ref{section5}, we review the core concepts of real-time search algorithm, which can be implemented without prior terrain information. Additionally, in the case study, we have advanced the algorithm's capability from serving a single user with one UAV to serving multiple users. 
\item In Sec.~\ref{section6}, we discuss several factors that can not be ignored in terrain-based UAV deployment but have received limited attention. 
\end{itemize}
}

\section{Classifications of Existing Literature}\label{section2}
In this subsection, we classify the existing terrain-based UAV positioning studies from different perspectives. Table~\ref{table1} displays the categories to which the existing studies belong under different classifications. The classification based on terrain information completeness determined the main framework of this paper. The correspondence between this classification and the other classification approaches is also a key focus in this subsection.

\subsection{Terrain Information Completeness}
As mentioned, terrain information completeness is classified into complete, incomplete, and no terrain information. 

\subsubsection{Complete Terrain Information}
Complete terrain information means the available terrain information enables the terrain construction to a certain degree of granularity \cite{gesbert2022uav}. This information should at least include the positions, heights, and shapes of obstacles such as buildings. In such situations, the most critical issue is how to acquire and periodically update three-dimensional (3D) terrain maps at a relatively lower cost. 

\subsubsection{Incomplete Terrain Information}
Due to the high cost of getting and updating complete 3D maps, a compromise solution is to record the features of occlusions with a few parameters. For instance, we can characterize the building density by the ratio of the building footprint area to the total area of the region \cite{al2014modeling}. Undoubtedly, since the available terrain information is compressed as characteristic parameters, UAVs are unable to accurately verify occlusion, resulting in a coarse-grained deployment. In this case, the most critical task is to map these characteristic parameters onto the blockage probability of the UAV-user link. 

\subsubsection{No Terrain Information}
No terrain information refers to the condition that prior information can not provide direct assistance or does not have a direct impact on the UAV deployment. Without relying on prior terrain information, UAVs have to collect terrain information during networking to achieve terrain-based deployment. Due to the limitations on power, UAVs have a restricted search trajectory length during real-time networking. Finding the optimal deployment position that avoids obstacles' occlusion has become the most critical issue, achieved through searching the shortest possible distance \cite{zhang2020radio}.

\subsection{Network Topology}
In terms of network topology, there are three types of transmission modes: single-hop transmission, relay transmission, and one-to-many transmission. 

\subsubsection{Single-Hop}
Single-hop communication is the fundamental communication structural unit, thus it has been mentioned in all three cases of terrain information completeness \cite{esrafilian2018learning,armeniakos2020sir, zhang2020radio}. The most extensively discussed aspect of single-hop communication is the issue of blockage detection. This involves determining the blockage probability based on the relative positions of the UAV and the user, as well as designing a moving trajectory when blocked to re-establish the LoS link.

\subsubsection{Relay}
In the no terrain information scenario, researchers have simplified the complex study of one-to-many transmission by focusing on relay transmission scenarios to optimize UAV deployment \cite{chen2019efficient}. Due to the absence of a direct LoS link between the transmitter and receiver, they need to perform relay transmission through a UAV located in their common LoS area. In relay transmission, the key issue is how to find the common LoS region and simultaneously optimize the performance of both links.

\subsubsection{One-to-Many}
One-to-many communication is suitable for scenarios with prior terrain information. When complete terrain information is available, UAVs achieve real-time deployment through fine-grained occlusion assessment \cite{huang2021navigating,mou2021deep}. However, in the absence of complete terrain information, UAVs can only resort to statistical models for coarse-grained deployment \cite{alzenad20173,he2018towards}.


\subsection{Stage}
Based on whether UAVs are initiating network services, the acquisition and utilization of terrain information by UAVs can be divided into two stages: real-time and offline. 

\subsubsection{Real-Time}
In the no terrain information scenario, UAVs acquire terrain information and deploy based on terrain after starting service \cite{zhang2020radio}. Therefore, no terrain information corresponds to the real-time stage.

\subsubsection{Offline}
Some tools can take terrain information as inputs, mapping them to target performance metrics, such as coverage probability and data rate \cite{wang2022ultra,wang2022stochastic}. As the entire process can be completed before networking, the deployment in the incomplete terrain information scenario is done in the offline stage.

\subsubsection{Hybrid}
As for UAV deployment with complete terrain, it is primarily divided into the offline stage of map construction and the real-time stage involving UAV flight trajectory design with the assistance of the map \cite{huang2021navigating}. Therefore, it is a deployment method with a hybrid blend of offline and real-time stages.

\section{Complete Terrain Information}\label{section3}
{  In this section, we review two crucial aspects in the context of complete terrain information: the UAV-aided terrain construction method and the trajectory design method for UAVs in dynamic scenarios. Then, a case study is proposed as an example of the interaction of two aspects of research content.}

\subsection{UAV-Aided Terrain Construction}
The UAV-aided terrain construction method is one of the key techniques for obtaining complete terrain information \cite{esrafilian20173d}. This method creates a map that records the signal strength at different locations and infers the position and shape of obstacles by map. Its principle is verifying blockage based on the difference in signal strength between LoS and non-line-of-sight (NLoS) links.

\par
{  According to the air-to-ground channel model proposed in \cite{alzenad2019coverage}, the received power $S_{Q}$ can be expressed as}
\begin{equation}\label{s_Q}
    S_{Q} =
     \zeta \, G_{Q} \,\eta_{Q}\, d^{-\alpha_{Q}}, 
\end{equation}
where $d$ is the Euclidean distance between the user and the UAV. $G_Q$ denotes the small-scale fading of the channel, which follows Nakagami-$m$ fading with scale parameter $m_Q$ \cite{wackerly2014mathematical}. The meanings and the recommended values of the remaining parameters are provided in Table~\ref{table2}. $Q$ in (\ref{s_Q}) is substituted by LoS if a LoS UAV-user link is established, otherwise $Q$ is replaced with NLoS.

\begin{table}[t]
\centering
\caption{Parameters of the air-to-ground channel model \cite{alzenad2019coverage}}
\label{table2}
\begin{tabular}{|c|c|cc|}
\hline
Meaning & Notation & \multicolumn{1}{c|}{LoS}     & NLoS     \\ \hline
Average additional loss & $\eta_{Q}$        & \multicolumn{1}{c|}{$-35$~dB} & $-48$~dB \\ \hline
Path-loss exponent  & $ \alpha_{Q} $      & \multicolumn{1}{c|}{$2$}       & $2.3$       \\ \hline
Scale parameter & $ m_{Q} $ & \multicolumn{1}{c|}{1}    & $2$       \\ \hline
Transmission power of UAV &  $\zeta$ & \multicolumn{2}{c|}{$30$~dBm} \\ \hline
\end{tabular}
\end{table}

\begin{figure}[th]
	\centering
    \includegraphics[width = 0.8 \linewidth]{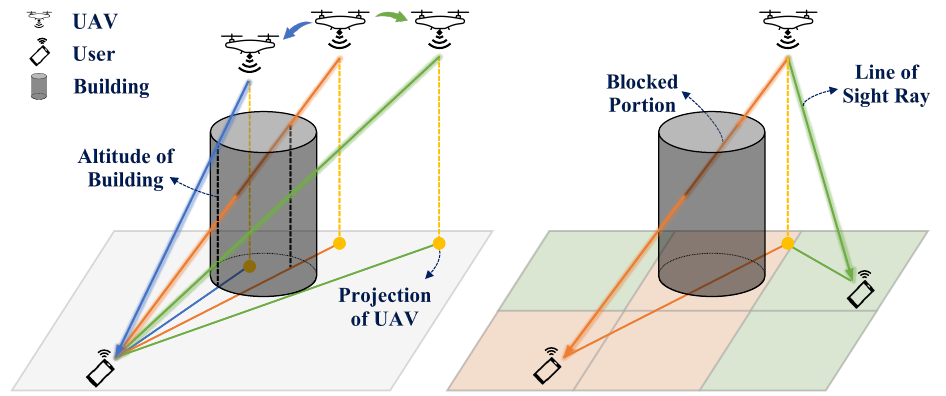}
	\caption{ Schematic diagram of UAV--aided terrain construction principles.}
	\label{figure9}
\end{figure}
The left part of Fig.~\ref{figure9} shows the principle of UAV-aided terrain construction. The link between the middle UAV and the ground user (indicated by the orange line) is blocked by a building, resulting in the user's received power being $S_{\mathrm{NLoS}}$. When the UAV approaches the user and crosses over the upper contour of the building shadow, the UAV-user link (the blue line) becomes LoS, and the user's received power is $S_{\mathrm{LoS}}$. Typically, the received signal power exhibits continuous variations with the continuous movement of the UAV. However, when crossing over the shadow contour of the building, according to (\ref{s_Q}), the received signal power generally experiences a drastic change exceeding $20$~dB. Therefore, the upper contour of the building should be located on the blue line. Similarly, the UAV can detect the right contour of the building shadow by shifting to the right.

\par
Furthermore, there are two remarks worth mentioning. Firstly, to accurately determine the building's position, height, and shape, it often requires the participation of multiple users (or ground signal collectors) in the construction process. In fact, with a larger number of users, the construction can process faster. The right of Fig.~\ref{figure9} provides an example of multi-user-assisted construction, where orange and green squares represent blocked and unblocked areas respectively. Secondly, the segmented propagation model is proposed to describe different degrees of attenuation caused by different types of obstacles, such as solid obstacles (e.g., buildings) and light obstacles (e.g., vegetation blockage). With this model, a greater variety of obstacles can be constructed with more precision and detail \cite{wu2020learning}.


\subsection{Trajectory Design in Dynamic Scenario}
The design of UAV trajectories is a critical topic in many applications, such as surveying, searching, and monitoring, as it directly influences the effectiveness and performance of UAV mission execution \cite{huang2021navigating,esrafilian2020three}. So far, with the robust computational power of computers, the trajectory planning of UAVs in static scenarios can be simulated through stochastic algorithms in the offline stage. Terrain-based UAV trajectory design typically operates within a small urban area, and in static scenarios with complete terrain information, it can even be accomplished through exhaustive search. Therefore, trajectory design in dynamic scenarios holds greater significance and poses more challenges.

\par
In dynamic scenarios, UAVs need to adjust their deployment positions based on the real-time locations of users. At this point, if the user's movement direction is entirely random, the UAV can not determine its deployment position for the next moment. Therefore, authors in \cite{huang2021navigating} proposed a scenario in which the UAV monitors the parade crowd moving according to a specific trajectory. Given that complete terrain information is available, UAVs attempt to get closer to the crowd to achieve better monitoring effectiveness while avoiding obstruction from buildings.

\subsection{Case Study: UAV Tracking During the Parade}\label{subsection3-3}
\subsubsection{Research Objective}
{ 
In existing studies, the topics of UAV-aided terrain construction and trajectory design in dynamic scenario have been discussed separately. However, the two should be closely connected. On the one hand, evaluating the effectiveness of terrain construction in real-world application scenarios is the most direct approach. On the other hand, assuming the availability of complete terrain information for dynamic UAV tracking does not align with practical scenarios in many cases. Therefore, the most important objective of this case study is to explore the interaction between terrain construction and dynamic UAV tracking. Given the known parade route, UAVs can primarily focus on terrain construction in the vicinity of the route to reduce the amount of terrain information collected. Conversely, the impact of errors in terrain construction on UAV tracking can also be studied.
}

\par
In this case study, we first utilize UAVs for terrain construction, followed by real-time UAV tracking of parade crowds. The constructed terrain assists the UAVs in avoiding blocking from buildings during tracking trajectory design. The algorithm provided in \cite{huang2021navigating} has been demonstrated to effectively track the crowd while avoiding blockage from buildings. However, the scenario is simplified, with users' potential positions restricted to several discrete points along the parade route and the shadow areas of buildings are represented by regular polygon topologies. Therefore, we test the effectiveness of the proposed algorithm in a more general scenario.

\subsubsection{UAV-Aided Terrain Construction}

\begin{figure}[th]
	\centering
	\includegraphics[width = 0.99\linewidth]{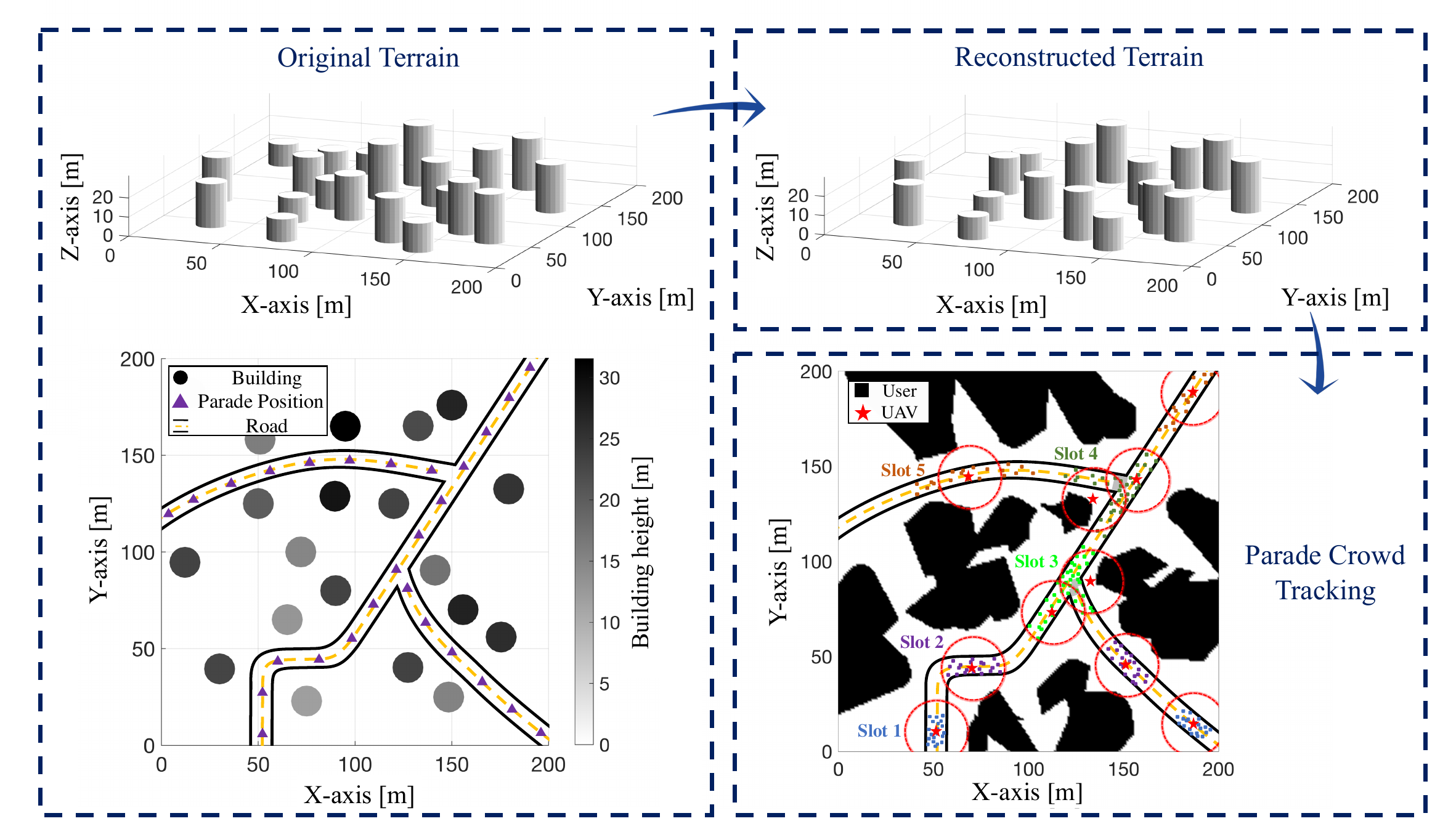}
	\caption{UAV-aided terrain construction and parade crowd tracking.}
	\label{fig:parade}
\end{figure}

The left upper and lower parts of Fig.~\ref{fig:parade} show the 3D and planar city topology, respectively. The buildings are distributed in a square area of $1$~km$\times$$1$~km. 
To avoid the effect of the building orientation, we refer to the recommendations in \cite{al2020probability} and model the building as a cylinder with a radius of $8$~m. The center of the cylindrical building follows a homogeneous Poisson point process (PPP) with a density of $500~ {\mathrm{buildings}}/{\mathrm{km}}^2$. The height of the building obeys log-normal distribution with parameters $3$ and $0.4$, which are the mean and standard deviation of the height’s logarithm, respectively\footnote{{  Note that all of the case studies are performed by MATLAB. The buildings, UAVs, and users' location generations and sampling procedures are implemented through self-written scripts (with "Editor" in MATLAB).}}.

\par
We only deploy ground signal receivers near the parade route, and UAVs also hover only in these areas, constructing terrain based on the algorithm proposed in \cite{esrafilian20173d}. As shown in the upper right part of Fig.~\ref{fig:parade}, the constructed buildings have an acceptable deviation from the original terrain in the vicinity of the route. However, several buildings far from the route have a significant error in terms of altitude, even some buildings have not been successfully constructed.

\subsubsection{Parade Crowd Tracking}
As shown in the right lower part of Fig.~\ref{fig:parade}, two sets of crowds are moving at an average pace of $50$~m/slot along the predefined parade trajectory on the road. The shaded part provides the blocked area when the UAV is at $30$~m height. Each crowd is tracked by one UAV. The positions of each user in the two crowds at different time slots are marked in the right lower part of Fig.~\ref{fig:parade}. The users in the crowd are generated near the preset central position and follow PPP at slot $1$. One crowd departs from the bottom right and exits on the left, while another crowd departs from the bottom left and exits on the top right. There is a segment of overlapping routes between two intersections at slots $3$ and $4$, leading to a certain degree of mixing of the crowds. Furthermore, the coverage is significantly influenced by terrain blockage at slots $3$ and $4$. With the help of the constructed terrain, the flight path of the UAV can more accurately avoid the blockage of the building and adapt to the more complex movement trajectory during these challenging slots for UAV tracking.

\section{Incomplete Terrain Information}\label{section4}
This section reviews the A2GLPM, where the LoS probability is the probability of establishing the LoS link between the transmitter and the receiver. After that, we summarize how existing studies deploy UAV networks and analyze network performance based on this model. Considering that the accuracy of the model has not been fully validated, we propose a case study to analyze the accuracy of the model itself and the accuracy of network performance analysis based on this model.

\subsection{LoS Probability Model}\label{subsection4-1}
The International Telecommunication Union (ITU) has proposed a standardized model for urban areas based on three simple terrain feature parameters that can describe the general geometric statistics of cities \cite{ITU-R}:
\begin{itemize}
    \item $\kappa$ represents the ratio of the buildings' footprint area to the total area (dimensionless).
    \item $\iota$ represents the average number of buildings per unit area (buildings$/$km$^2$).
    \item $\omega$ is the scale parameter of Rayleigh distribution that describes the height of the building (dimensionless). 
\end{itemize}

\begin{figure}[ht]
	\centering
	\includegraphics[width = 0.65\linewidth]{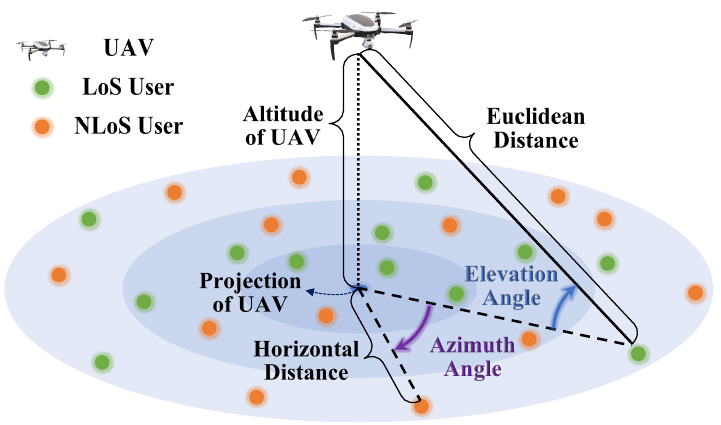}
	\caption{ { Schematic diagram of user distribution and UAV-user relative position. }}
	\label{fig:7}
\end{figure}

\par
Based on these three parameters, the probability of the UAV and user establishing an LoS link $P_{\rm{LoS}}(\theta)$ is given by the A2GLPM \cite{al2014optimal},
\begin{equation}\label{PLoS}
\begin{split}
    P_{\rm{LoS}}(\theta) = \frac{1}{1 + a\exp \left( { - b\left( \theta - a \right)} \right)},
\end{split}
\end{equation}
where $\theta = \arctan \left( {{h}/{\sqrt{d^2-h^2}}} \right)$ is the elevation angle, $d$ is the Euclidean distance between the UAV and user, $h$ is the altitude of UAV. An example of the elevation angle is given in Fig.~\ref{fig:7}. $a$ and $b$ can be presented by the terrain feature parameters:
\begin{equation}\label{a and b}
    Q = \sum_{j=0}^3 \sum_{i=0}^{3-j} C_{i,j} \left(\kappa \iota \right)^i \omega^j,
\end{equation}
where $Q \in \{a,b\}$, $C_{i,j}$ are polynomial coefficients. The values of $C_{i,j}$ can refer to Table \uppercase\expandafter{\romannumeral1} and Table \uppercase\expandafter{\romannumeral2} in \cite{al2014optimal}.


As a result, the A2GLPM maps the terrain feature parameters to the blockage probability, given the relative positions of the UAV and user.

\subsection{Stochastic Geometry (SG) Analytical Framework} \label{subsection4-2}
The authors in \cite{alzenad2019coverage} integrated the A2GLPM into the SG analytical framework, laying the foundation for subsequent extensive research in this field. From the perspective of a typical UAV, users can be categorized into LoS users and NLoS users. Specifically, all of the users are assumed to follow a homogeneous PPP with density $\lambda$. Then, the PPP is divided into two sub-PPPs:
\begin{itemize}
    \item An LoS-PPP with density $\lambda P_{\mathrm{LoS}} (\theta)$ records the positions of users that are not blocked (referred to as LoS users). $P_{\mathrm{LoS}}(\theta)$ is defined in (\ref{PLoS}). 
    \item An NLoS-PPP with density $\lambda \left(1-P_{\mathrm{LoS}}(\theta) \right)$ records the positions of users that are blocked (referred to as NLoS users).
\end{itemize}
As shown in Fig.~\ref{fig:7}, when the altitude of the UAV is fixed, the density of LoS users decreases with increasing distance to the UAV.

\par
Then, we take the downlink coverage probability from the perspective of a typical user as an example to illustrate how terrain information influences the SG analyzing process. The typical user is served by the UAV that provides the maximum average received power, while other UAVs are interfering UAVs. The interference power is the cumulative received power at the user side, attributed to interfering UAVs. The coverage probability is defined as the probability that the received signal-to-interference plus noise ratio is larger than a given coverage threshold \cite{lou2023coverage}. From this, it can be observed that coverage probability is closely linked to the distances between users and LoS and NLoS UAVs, and to a large extent, it depends on the accuracy of the A2GLPM.

\par
With the SG analytical framework, coverage probability can be expressed as a function of terrain feature parameters and network distribution parameters, such as UAV deployment density and altitude. Therefore, the derived expression for coverage probability can not only be utilized for performance estimation but also serve as a tool for optimizing the deployment of UAV networks. However, due to incomplete terrain information, the deployment is coarse-grained. Compared to existing deployment algorithms in the complete terrain scenarios, the SG analytical framework also possesses its unique advantage, that is, the capability for interference analysis \cite{alzenad2019coverage}.

\subsection{Case Study: Accuracy of the Model}\label{subsection4-3}
\subsubsection{Research Objective}
The previous subsection emphasized the importance of model accuracy in performance analysis. Therefore, this subsection has three objectives regarding accuracy verification. First of all, to achieve a concise expression, A2GLPM neglects the correlation of user positions. In other words, the probability of two adjacent users being blocked remains mutually independent in A2GLPM, while two adjacent users are likely to be blocked simultaneously in reality. Therefore, we hope to investigate whether the correlation between user positions can be neglected in the blockage verification. Secondly, the A2GLPM is presented as a modified Sigmoid function containing two degrees of freedom, namely, $a$ and $b$. An exploratory issue is whether other commonly applied functions involving two degrees of freedom can provide a more accurate description of the relationship between elevation angle and LoS probability. Finally, we aim to find the difference between the approximate coverage probability results obtained through A2GLPM for LoS probability and the accurate coverage probability obtained through simulation with precise blockage verification.

\subsubsection{Model Accuracy Analysis}
\begin{figure}[ht]
	\centering
	\includegraphics[width = 0.6\linewidth]{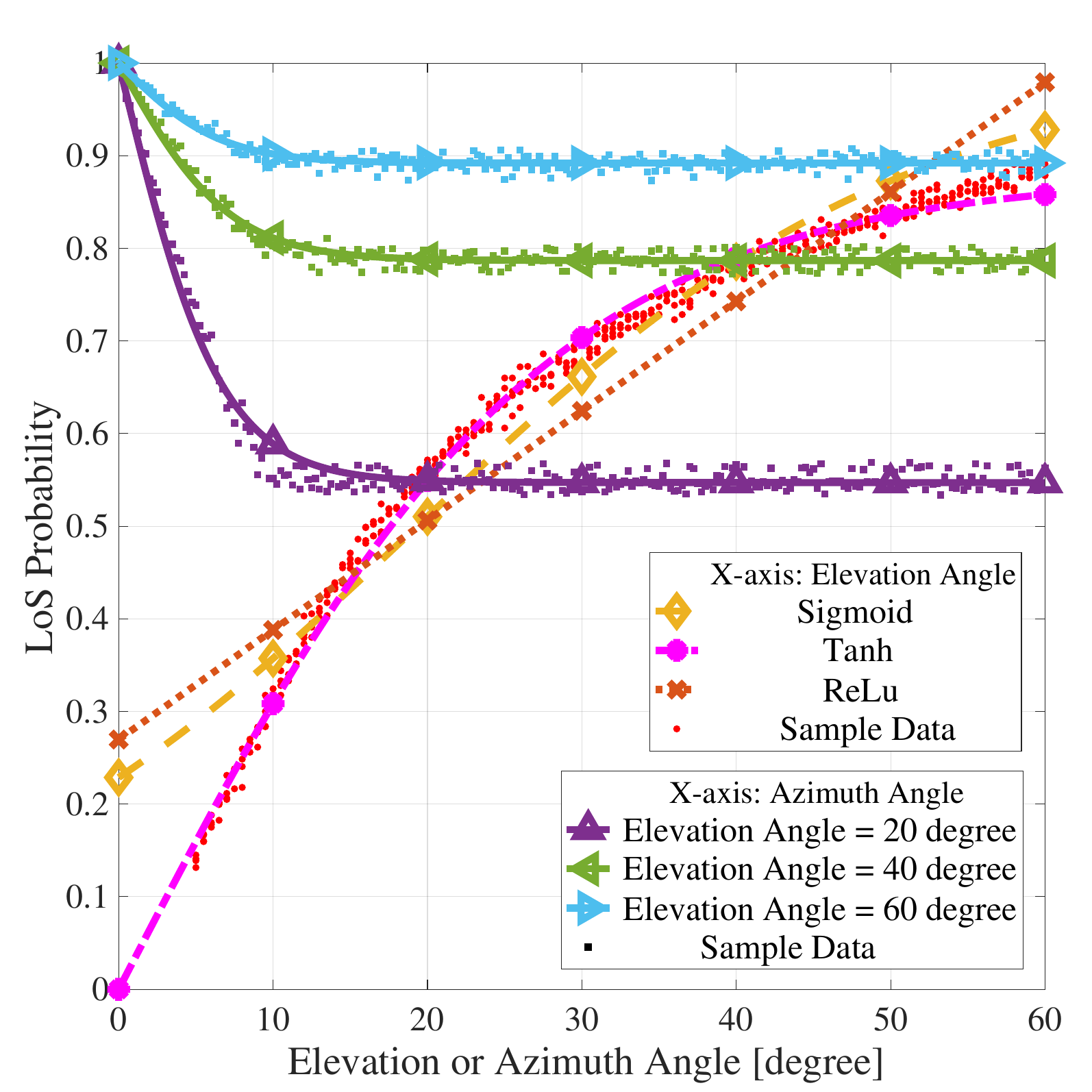}
	\caption{LoS probability at different elevation and azimuth angles.}
	\label{fig:Angle}
\end{figure}

In this case study, UAVs and users are distributed in a square area of $1$~km$\times$$1$~km, forming two homogeneous PPPs. The density of users is fixed as $1000 ~ {\mathrm{users}}/{\mathrm{km}}^2$. The height of UAVs is $80$~m. The distribution of buildings is the same as in Sec.~III-\ref{subsection3-3} In Fig.~\ref{fig:Angle}, we plot the sample data obtained by simulation with dots. For a fixed elevation angle, the LoS probability is calculated by the proportion of the number of samples of UAV-user links not blocked by cylindrical buildings in the total number of samples.

\par
The relationship between elevation angle and LoS probability is fitted with three different functions:
\begin{itemize}
    \item Modified Sigmoid function: $[a,b] = [2.8240, 0.0628]$,
    \begin{equation}
    P_{\mathrm{LoS}}^{\mathrm{Sig}} (\theta) = \frac{1}{1+{a\exp(-b(\theta-a))}}.
    \end{equation}
    \item Modified Tanh function: $[a,b] = [0.8794, 2.0984]$,
    \begin{equation}
    P_{\mathrm{LoS}}^{\mathrm{Tanh}} (\theta) = a \, \frac{ \exp (2b\theta)-1}{\exp (2b\theta)+1}.
    \end{equation}
    \item Modified ReLu function: $[a,b] = [0.6775, 0.2697]$,
    \begin{equation}
    P_{\mathrm{LoS}}^{\mathrm{ReLu}} (\theta) = \max\{0,a\theta+b\}.
    \end{equation}
\end{itemize}
The parameters $[a,b]$ are obtained from the minimum mean square error (MSE) between the sample data and the corresponding value of the fitted curve. The MSEs for Sigmoid, Tanh and ReLu functions are $\{ 2.12, 0.25, 4.26 \} \times 10^{-3}$ respectively. {  The accuracy of the fitting by the three functions can also be visually observed through Fig.~\ref{fig:Angle}. The above results indicate that the improved Sigmoid function in \cite{al2014modeling} can accurately describe the relationship between LoS probability and elevation angle. However, there might be better alternatives, such as the improved Tanh function. 
}

\par
Next, the correlation between user positions is measured by the azimuth angle. As drawn in Fig.~\ref{fig:7} the azimuth angle takes the projection of the UAV on the ground as the endpoint and the rays from the projections to two users (denoted as user A and user B) as the edges. { When calculating the LoS probability for azimuth angle in Fig.~\ref{fig:Angle}, we verify whether the UAV-user B link is blocked, given that a LoS link is established between UAV and user A. At a fixed elevation angle, the LoS probability is independent of the azimuth angle as long as the azimuth is greater than a threshold, which is around $10$ degrees in this case, as shown in Fig.~\ref{fig:Angle}.} The threshold of independence is related to the footprint area of buildings. Furthermore, under a fixed elevation angle $\theta_0$, the LoS probability for azimuth angle converges to $P_{\mathrm{LoS}} (\theta_0)$ as the azimuth angle increases.

\subsubsection{Network Performance Accuracy Analysis}
\begin{figure}[ht]
	\centering
	\includegraphics[width = 0.6\linewidth]{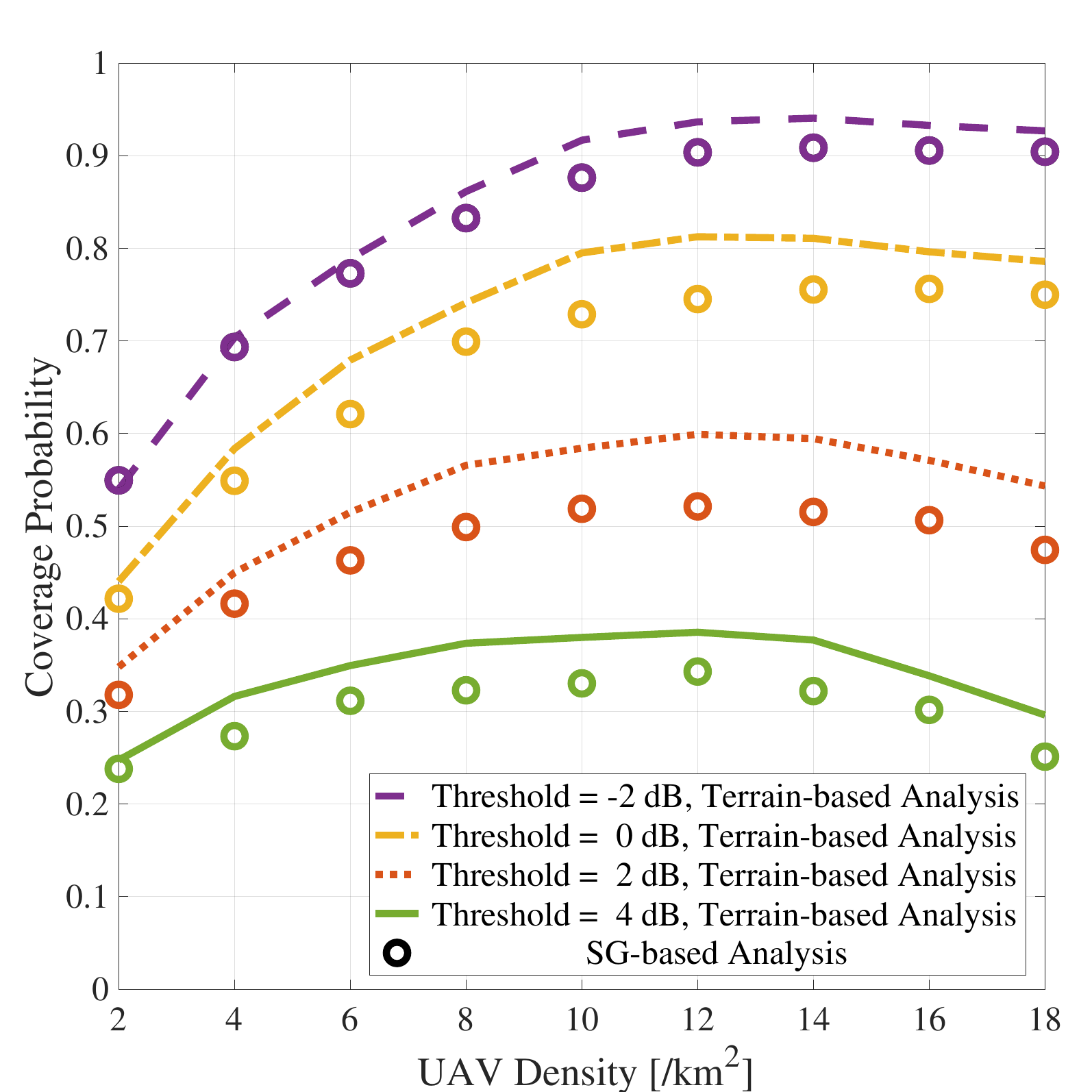}
	\caption{Coverage probability with different UAV densities.}
	\label{fig:Coverage}
\end{figure}

Now, we identify the variance between the approximate coverage probability results obtained by the SG framework and the accurate coverage probability obtained through simulation with accurate blockage verification. Unless otherwise stated, channel parameters are set to their default values given in Table~\ref{table2}. The environmental noise power is $\sigma^2 = -90$~dBm. The points labeled as "SG-based Analysis" in Fig.~\ref{fig:Coverage} are obtained by Theorem~2 in  \cite{alzenad2019coverage}. The A2GLPM follows the modified Sigmoid function with terrain related parameters $[a,b] = [2.8240, 0.0628]$. For lines labeled as "Terrain-based Analysis" in Fig.~\ref{fig:Coverage}, complete terrain information is available for UAVs. Blockage verification is no longer based on any statistical model but is determined through simulation.

Fig.~\ref{fig:Coverage} shows that the coverage probability obtained under the SG framework has about a $9\%$ relative deviation compared to the accurate one. Given that the SG-based approach relies on only two parameters to incorporate terrain information, the relative deviation is acceptable, especially when the coverage probability is high. Another interesting observation that can be made from Fig.~\ref{fig:Coverage} is that the coverage probability estimated through the SG framework is consistently slightly smaller than the accurate values. {  The primary reason for this phenomenon is that the A2GLPM underestimates the LoS probability for associated UAV or overestimates the LoS probability for interfering UAVs.}

\par
Furthermore, this case study validates the feasibility of optimizing UAV deployment density through SG-based analytical results. Fig.~\ref{fig:Coverage} shows the UAVs' density that maximizes the coverage probability is around $12$~UAVs$/$km$^2$, which is independent of the coverage threshold. When the deployment density of UAVs is low, users may be at a considerable distance from the serving UAV, and there is a possibility of no UAV within the line of sight being available. Conversely, if the deployment density is too high, interference becomes a primary factor limiting coverage probability. It is evident that the optimization of network-level parameters, such as UAV density, relies on the unique interference analysis capability of the SG framework. The optimal value is challenging to attain through the design of UAV tracking or search algorithms mentioned earlier.


\section{No Terrain Information}\label{section5}
This section centers on the exploration of the real-time search algorithm. A detailed overview of the algorithm's objectives, principles, and core concepts is provided. Additionally, the specific execution process is illustrated through heat maps. Considering that the existing methods only apply to the relay scenario, we extended the algorithm to one-to-many scenario. Since the real-time search does not require prior terrain information, it is limited by many factors and requires trade-offs between practicability and performance.

\subsection{Real-time Search Method}\label{subsection5-1}
In simple terms, the real-time search method is an online realization of UAV-assisted terrain construction. Real-time search algorithms are typically designed to address a class of max-min problems in the field of wireless communication. The application scenario in \cite{chen2019efficient} is taken as an example to illustrate the optimization problem. In this scenario, the UAV serves as a relay between a user and an aerial base station (ABS). The optimization objective is to identify an optimal UAV deployment position that maximizes the smaller of the two SNRs: the average SNR of the ABS-UAV link and the average SNR of the UAV-user link. It can be mathematically represented as a special case of the following optimization problem when $N=2$:
\begin{equation}\label{optimization}
	\underset{x \in \mathcal{X}}{\arg\max} \ \min \left\{ \frac{ \mathbb{E}_G [S_{Q,1}] }{\sigma^2}, \frac{ \mathbb{E}_G [S_{Q,2}] }{\sigma^2}, \dots , \frac{ \mathbb{E}_G [S_{Q,N}] }{\sigma^2} \right\}
\end{equation}
$N$ represents the number of devices simultaneously served by the UAV. $x$ denotes the position of the UAV, and $\mathcal{X}$ is the set recording the locations that are allowed for deployment. $S_{Q,1}$ and $S_{Q,2}$ are receiver power of ABS-UAV link and UAV-user link, where $Q=\{ \mathrm{LoS}, \mathrm{NLoS}\}$. Due to the fact that received power is not only dependent on the distance between the UAV and the serviced device but also on whether the link is blocked. Since the terrain is complex and irregular, this problem is non-convex and can only be solved through algorithmic search.

\par
Although the real-time search algorithm can be considered as a specific type of UAV-assisted terrain construction method, there are still distinct differences between the two \cite{lou2024terrain}. Firstly, the search trajectory of real-time search algorithms should be linear, because the limited power supply of UAVs makes it challenging to support search trajectories with quadratic or higher complexity \cite{kishk2020aerial}. Linearity is typically in reference to the distance between the two furthest serviced users. Due to strict limitations on the search trajectory length, the search direction at each iteration needs to be carefully designed based on terrain information gathered through previous searches. In addition, real-time search involves the concept of granularity, which can be understood as the step size of the UAV at each iteration. Due to the need for the UAV to initiate communication with users each time it reaches a new search position to assess whether the link is blocked, larger granularity allows the UAV to search for potential optimal positions more quickly. However, it also increases the probability of missing the optimal position.

\subsection{Case Study: The Extension of Real-time Search} \label{subsection5-2}
\subsubsection{Research Objective}
In this section, we first illustrate the core idea of the algorithm through a relatively simple example to provide an intuitive understanding. Additionally, we will clarify that the algorithm's search trajectory is linear, and how it is designed to search for potential optimal positions. Secondly, since the previous study focuses on a relatively simple relay scenario, and one UAV only serves one user. Therefore, the case study aims to generalize the existing method to a one-to-many scenario. Finally, we will compare the average coverage performance under different levels of terrain completeness.

\subsubsection{Method Principle}

\begin{figure}[ht]
	\centering
	\includegraphics[width = 0.99\linewidth]{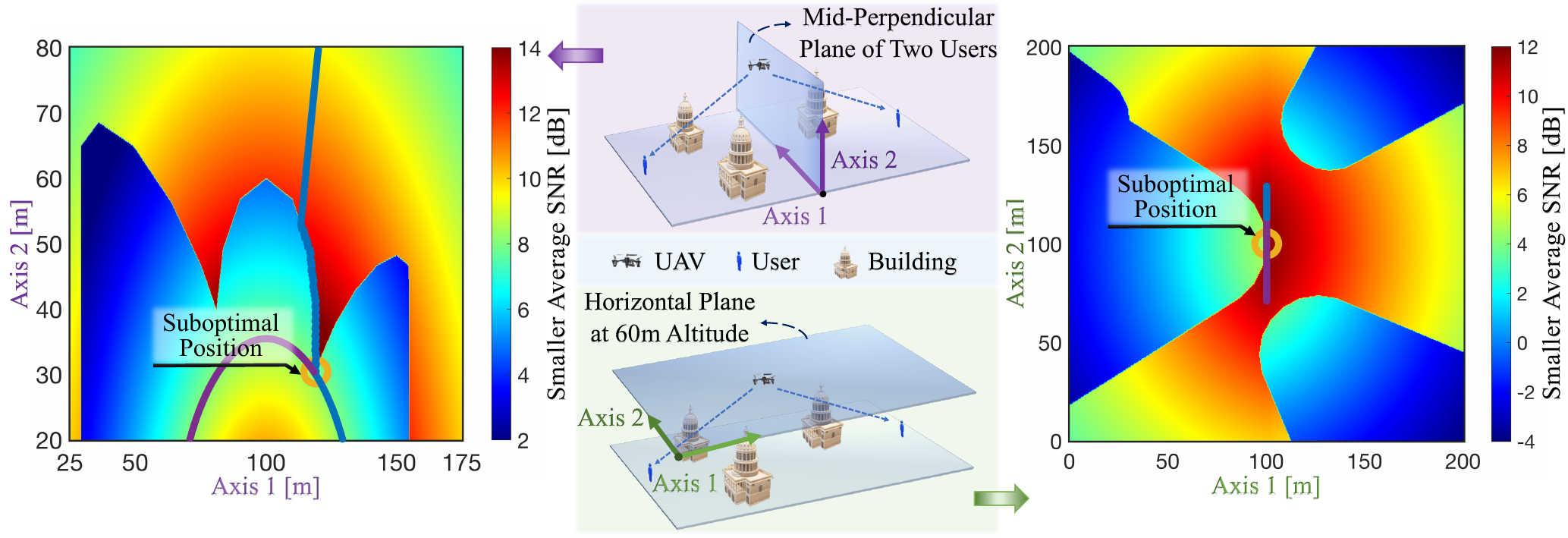}
	\caption{An example of real-time search.}
	\label{fig:heatmap}
\end{figure}

In this case study, the granularity of UAV search is $0.2$~m, and the coverage threshold is set to $4$~dB. The optimization objective remains as specified in (\ref{optimization}). To facilitate the understanding of algorithmic concepts, we start from a local area on the map and discuss a relatively simple scenario. The scenario involves one UAV, two users, and three buildings, which is shown in the middle part of Fig.~\ref{fig:heatmap}. The real-time search algorithm is based on two core ideas:
\begin{itemize}
    \item $(c_1)$: When the UAV is not blocked, it moves in the direction where there is a steep increase in the average SNR.
    \item $(c_2)$: When the UAV is blocked, the UAV moves along a trajectory where the average SNR does not change to avoid missing the next LoS region. 
\end{itemize}

\par
The objective of the max-min problem is to emphasize that UAVs can provide equitable service quality for two users. Therefore, a significant portion of the UAV's search trajectory is likely to be along a plane where the optimal deployment position is equidistant to both users, i.e., along the mid-perpendicular plane. In the left part of Fig.~\ref{fig:heatmap}, we plot a heat map to show the smaller average SNR on the mid-perpendicular plane of two users. In this heat map, the contour of buildings' shadows is quite clear due to the intense changes in SNR. 

\par
The UAV starts from above, initially descending rapidly based on the first core idea $(c_1)$ until it reaches the contour of the building's shadow. Then, part of the trajectory overlaps with the contour because the UAV repeatedly enters the NLoS region of one user according to $(c_1)$ and then comes back to the common LoS area according to $(c_2)$ in a bent route. Finally, after reaching the suboptimal position, the UAV follows a left search trajectory (purple curve) and a right search trajectory (blue curve) where the average SNR remains constant on these trajectories according to $(c_2)$. The marked suboptimal position may not necessarily be optimal because the optimal position may not lie within the mid-perpendicular plane{\footnote{In fact, a suboptimal position with an elevation angle to the users of less than 45 degrees is a sufficient but not necessary condition for it to be the optimal relay position.}}. As shown in the right part of Fig.~\ref{fig:heatmap}, the trajectory from the perspective of the horizontal plane implies the searching trajectory is linear.

\subsubsection{Generalized Algorithm}
Next, we generalize the proposed algorithm into a multi-UAV serving multi-user scenario. The buildings follow the same distribution as in Sec.~III-\ref{subsection3-3} The user distribution, UAV distribution, and channel model are the same as in Sec.~IV-\ref{subsection4-3}. The density of UAV is set to optimal $12$~UAVs$/$km$^2$ one obtained in Sec.~IV-\ref{subsection4-3}. 

\par
The core idea of the generalization is simply described as follows. The UAV focuses on the two most distant associated users to perform the proposed algorithm. Starting from a random position, the UAV takes the shortest path to reach the mid-perpendicular plane and then executes the proposed real-time search algorithm. In this process, the two users farthest from the UAV may undergo changes, thereby causing the algorithm to initiate a repetition of the initial steps. The UAV stops when the maximum search length is reached, or the algorithm finishes.

\begin{figure}[ht]
	\centering
	\includegraphics[width = 0.6\linewidth]{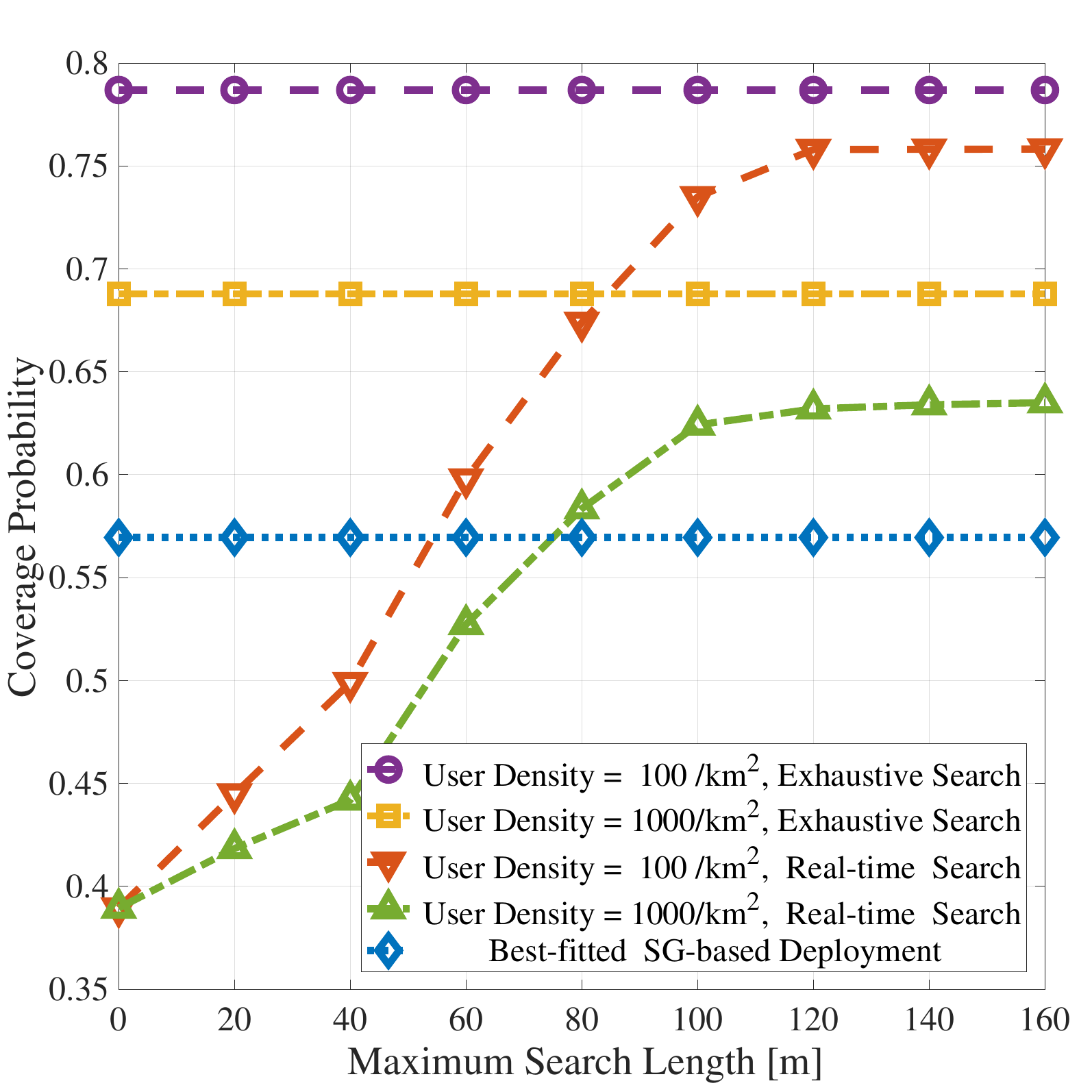}
	\caption{The influence of search length on the performance of real-time search.}
	\label{fig:Search}
\end{figure}

\par
In Fig.~\ref{fig:Search}, we compare the coverage performance in scenarios with complete, incomplete, and no terrain information. The 3D exhaustive search is executed in the scenario with complete terrain information to serve as an upper bound. In the best-fitted SG-based deployment, UAVs follow a PPP at the optimal altitude that maximizes the coverage probability given the density is $12$~UAVs$/$km$^2$. This is considered as UAV deployment under incomplete terrain information. In no terrain information, once the search trajectory length reaches the preset upper limit ($x$-axis in Fig.~\ref{fig:Search}), the search stops, and the UAV is deployed to the position with the maximum coverage probability along the trajectory. The $y$-axis in Fig.~\ref{fig:Search}) represents the mean coverage probability for all users in the area. 

\par
As shown in Fig.~\ref{fig:Search}, with a smaller user density, the coverage performance of the real-time search algorithm is better and closer to the upper bound provided by exhaustive search. However, due to the limitations of a linear search trajectory, the UAV can not achieve complete terrain information through real-time searching. Therefore, regardless of the upper limit set for the search trajectory, the coverage performance of real-time search can never reach the upper bound provided by exhaustive search. The intersection point between the curve corresponding to the real-time search algorithm and the straight line corresponding to the SG-based deployment represents the moment when the terrain information obtained by the UAV through real-time search is equal to the terrain information contained by the terrain feature parameters.

\subsection{Trade-offs between Practicability and Performance}
Without prior terrain information required, the real-time search method can find the optimal position with a linear search distance. It is not difficult to imagine that the method suffers several drawbacks, resulting in trade-offs between practicability and performance.

\subsubsection{Optimality and Number of Served Users}
The real-time search method can ensure optimality when one UAV serves one user. When the UAV is serving two users, the optimality can only be satisfied under certain conditions. In one-to-many network topology, when a UAV approaches one user, the links with other users might be blocked, so achieving optimality is challenging.

\subsubsection{Update Frequency and Power Consumption}
Frequently updating real-time search will lead to a large amount of power consumption, probably resulting in the UAV's inability to maintain high-quality communication. However, a low updating frequency may also cause a degradation of communication quality as whether users enter the shadow of the building is unknown to the UAV.

\subsubsection{Search Length and Prior Terrain Information}
At each update, the determination of the starting search position partly depends on the prior terrain information. More terrain information often leads to a starting position closer to the potential optimal location, thus shortening the search length. Utilizing the terrain information recorded in the search process before updating can be an interesting research direction of real-time search methods.

\section{Further Discussions} \label{section6}
This section explores three overlooked factors in terrain-based UAV deployment that warrant attention.

\subsection{Charging}
As mentioned, power supply is an important factor affecting the hovering time and communication quality of UAVs. When charged by wireless laser power beams \cite{belmekki2022unleashing}, UAVs should also pay attention to whether the power supply equipment's link is blocked by obstacles, especially in high-rise urban regions. The establishment of aerial power stations with a higher altitude can improve charging efficiency.

\subsection{Backhaul}
In the one-to-many scenario, the backhaul link between the UAV and the base station from the core network bears the sum of data rates for all users \cite{lou2023haps}. For base stations with heights lower than buildings, UAVs can also ensure that the backhaul link is not blocked by employing real-time search methods. Additionally, UAVs should promptly reduce the distance to the base station when the backhaul link experiences a significant communication load.

\subsection{Electromagnetic field (EMF) exposure}
When lowering EMF exposure is emphasized, UAVs might be deployed below the optimal density to reduce EMF exposure from interference signals \cite{lou2021green}. Compared to offline deployment relying on terrain information, real-time search offers greater advantages. The optimal deployment positions identified through real-time search are characterized by lower altitudes, resulting in minimized interference to users.

\section{Conclusion}\label{section7}
The article classified and discussed terrain-based UAV network deployment methods from the perspective of terrain information completeness. It introduced the core algorithms and models involved, discussed their corresponding application scenarios, and provided case studies as examples. In the first case study, we demonstrated that UAVs could provide stable coverage for regularly moving users using existing local terrain construction methods with UAV sampling. In the second case study, a coarse-grained UAV deployment was performed using a simple set of parameters that characterized the terrain features. A stochastic geometry framework provided general analytical results for the aforementioned coarse-grained UAV networks. In the last case study, the UAV was able to avoid building blockage without prior terrain information through real-time linear-trajectory search.


\bibliographystyle{IEEEtran}
\bibliography{references_red}

\begin{thebibliography}{10}
\providecommand{\url}[1]{#1}
\csname url@samestyle\endcsname
\providecommand{\newblock}{\relax}
\providecommand{\bibinfo}[2]{#2}
\providecommand{\BIBentrySTDinterwordspacing}{\spaceskip=0pt\relax}
\providecommand{\BIBentryALTinterwordstretchfactor}{4}
\providecommand{\BIBentryALTinterwordspacing}{\spaceskip=\fontdimen2\font plus
\BIBentryALTinterwordstretchfactor\fontdimen3\font minus \fontdimen4\font\relax}
\providecommand{\BIBforeignlanguage}[2]{{%
\expandafter\ifx\csname l@#1\endcsname\relax
\typeout{** WARNING: IEEEtran.bst: No hyphenation pattern has been}%
\typeout{** loaded for the language `#1'. Using the pattern for}%
\typeout{** the default language instead.}%
\else
\language=\csname l@#1\endcsname
\fi
#2}}
\providecommand{\BIBdecl}{\relax}
\BIBdecl

\bibitem{ansari2021urban}
S.~Ansari, A.~Taha, K.~Dashtipour, Y.~Sambo, Q.~H. Abbasi, and M.~A. Imran, ``Urban air mobility—{A} {6G} use case?'' \emph{Frontiers in Communications and Networks}, vol.~2, p. 729767, 2021.

\bibitem{zaid2023aerial}
A.~A. Zaid, B.~E.~Y. Belmekki, and M.-S. Alouini, ``Aerial-aided {mmWave} {VANETs} using {NOMA}: Performance analysis, comparison, and insights,'' \emph{IEEE Transactions on Vehicular Technology}, 2023.

\bibitem{wang2023terrain}
R.~Wang, W.~U. Mondal, M.~A. Kishk, V.~Aggarwal, and M.-S. Alouini, ``Terrain-based coverage manifold estimation: Machine learning, stochastic geometry, or simulation?'' \emph{IEEE Open Journal of the Communications Society}, 2023.

\bibitem{huang2023system}
Q.~Huang, B.~E.~Y. Belmekki, A.~M. Eltawil, and M.-S. Alouini, ``System-level metrics for non-terrestrial networks under stochastic geometry framework,'' \emph{IEEE Communications Magazine}, 2024, accepted.

\bibitem{wang2024ultra}
R.~Wang, M.~A. Kishk, and M.-S. Alouini, ``Ultra reliable low latency routing in {LEO} satellite constellations: A stochastic geometry approach,'' \emph{IEEE Journal on Selected Areas in Communications}, 2024.

\bibitem{imran2019seamless}
M.~A. Imran, A.~Turkmen, M.~Ozturk, J.~Nadas, and Q.~H. Abbasi, ``Seamless indoor/outdoor coverage in 5{G},'' \emph{Wiley 5G Ref: The Essential 5G Reference Online}, pp. 1--23, 2019.

\bibitem{letaief2019roadmap}
K.~B. Letaief, W.~Chen, Y.~Shi, J.~Zhang, and Y.-J.~A. Zhang, ``The roadmap to {6G}: {AI} empowered wireless networks,'' \emph{IEEE Communications Magazine}, vol.~57, no.~8, pp. 84--90, 2019.

\bibitem{fan2022learning}
C.~Fan, C.~She, H.~Zhang, B.~Li, C.~Zhao, and D.~Niyato, ``Learning to optimize user association and spectrum allocation with partial observation in mm{W}ave-enabled {UAV} networks,'' \emph{IEEE Transactions on Wireless Communications}, vol.~21, no.~8, pp. 5873--5888, 2022.

\bibitem{shang2019unmanned}
B.~Shang, L.~Liu, J.~Ma, and P.~Fan, ``Unmanned aerial vehicle meets vehicle-to-everything in secure communications,'' \emph{IEEE Communications Magazine}, vol.~57, no.~10, pp. 98--103, 2019.

\bibitem{zaid2023evtol}
A.~A. Zaid, B.~E.~Y. Belmekki, and M.-S. Alouini, ``{eVTOL} communications and networking in {UAM}: Requirements, key enablers, and challenges,'' \emph{IEEE Communications Magazine}, vol.~61, no.~8, pp. 154--160, 2023.

\bibitem{fan2020robust}
C.~Fan, B.~Li, J.~Hou, Y.~Wu, W.~Guo, and C.~Zhao, ``Robust fuzzy learning for partially overlapping channels allocation in {UAV} communication networks,'' \emph{IEEE Transactions on Mobile Computing}, vol.~21, no.~4, pp. 1388--1401, 2020.

\bibitem{shang2018wireless}
B.~Shang, L.~Zhao, K.-C. Chen, and X.~Chu, ``Wireless-powered device-to-device-assisted offloading in cellular networks,'' \emph{IEEE Transactions on Green Communications and Networking}, vol.~2, no.~4, pp. 1012--1026, 2018.

\bibitem{gomez2013performance}
K.~Gomez, T.~Rasheed, L.~Reynaud, and S.~Kandeepan, ``On the performance of aerial {LTE} base-stations for public safety and emergency recovery,'' in \emph{2013 IEEE IEEE Open Journal of Vehicular Technology Workshops (GC Wkshps)}.\hskip 1em plus 0.5em minus 0.4em\relax IEEE, 2013, pp. 1391--1396.

\bibitem{khuwaja2018survey}
A.~A. Khuwaja, Y.~Chen, N.~Zhao, M.-S. Alouini, and P.~Dobbins, ``A survey of channel modeling for {UAV} communications,'' \emph{IEEE Communications Surveys \& Tutorials}, vol.~20, no.~4, pp. 2804--2821, 2018.

\bibitem{esrafilian2018learning}
O.~Esrafilian, R.~Gangula, and D.~Gesbert, ``Learning to communicate in {UAV}-aided wireless networks: Map-based approaches,'' \emph{IEEE Internet of Things Journal}, vol.~6, no.~2, pp. 1791--1802, 2018.

\bibitem{huang2021navigating}
H.~Huang and A.~V. Savkin, ``Navigating {UAV}s for optimal monitoring of groups of moving pedestrians or vehicles,'' \emph{Transactions on Vehicular Technology}, vol.~70, no.~4, pp. 3891--3896, 2021.

\bibitem{mou2021deep}
Z.~Mou, Y.~Zhang, F.~Gao, H.~Wang, T.~Zhang, and Z.~Han, ``Deep reinforcement learning based three-dimensional area coverage with {UAV} swarm,'' \emph{IEEE Journal on Selected Areas in Communications}, vol.~39, no.~10, pp. 3160--3176, 2021.

\bibitem{yi2022joint}
P.~Yi, L.~Zhu, L.~Zhu, Z.~Xiao, Z.~Han, and X.-G. Xia, ``Joint {3-D} positioning and power allocation for {UAV} relay aided by geographic information,'' \emph{IEEE Transactions on Wireless Communications}, vol.~21, no.~10, pp. 8148--8162, 2022.

\bibitem{esrafilian2020three}
O.~Esrafilian, R.~Gangula, and D.~Gesbert, ``Three-dimensional-map-based trajectory design in {UAV}-aided wireless localization systems,'' \emph{IEEE Internet of Things Journal}, vol.~8, no.~12, pp. 9894--9904, 2020.

\bibitem{al2020probability}
A.~Al-Hourani, ``On the probability of line-of-sight in urban environments,'' \emph{IEEE Wireless Communications Letters}, vol.~9, no.~8, pp. 1178--1181, 2020.

\bibitem{armeniakos2020sir}
C.~K. Armeniakos, P.~S. Bithas, and A.~G. Kanatas, ``{SIR} analysis in {3D UAV} networks: {A} stochastic geometry approach,'' \emph{IEEE Access}, vol.~8, pp. 204\,963--204\,973, 2020.

\bibitem{chen2023correlation}
{ L. Chen, W. Zhang, M. A. Kishk, and M.-S. Alouini}, ``Correlation of line-of-sight probabilities in aerial-terrestrial communications: Modeling, analysis, and application,'' \emph{IEEE Transactions on Vehicular Technology}, 2023\color{black}.

\bibitem{yin2023air}
{ D. Yin, X. Yang, H. Yu, S. Chen, and C. Wang}, ``An air-to-ground relay communication planning method for {UAVs} swarm applications,'' \emph{IEEE Transactions on Intelligent Vehicles}, 2023\color{black}.

\bibitem{wang2023resident}
R.~Wang, M.~A. Kishk, and M.-S. Alouini, ``Resident population density-inspired deployment of {K}-tier aerial cellular network,'' \emph{IEEE Transactions on Wireless Communications}, vol.~22, no.~11, pp. 7989--8002, 2023.

\bibitem{al2014optimal}
A.~Al-Hourani, S.~Kandeepan, and S.~Lardner, ``Optimal {LAP} altitude for maximum coverage,'' \emph{IEEE Wireless Communications Letters}, vol.~3, no.~6, pp. 569--572, 2014.

\bibitem{alzenad20173}
M.~Alzenad, A.~El-Keyi, F.~Lagum, and H.~Yanikomeroglu, ``3-{D} placement of an unmanned aerial vehicle base station ({UAV-BS}) for energy-efficient maximal coverage,'' \emph{IEEE Wireless Communications Letters}, vol.~6, no.~4, pp. 434--437, 2017.

\bibitem{he2018towards}
X.~He, W.~Yu, H.~Xu, J.~Lin, X.~Yang, C.~Lu, and X.~Fu, ``Towards {3D} deployment of {UAV} base stations in uneven terrain,'' in \emph{27th international conference on computer communication and networks (ICCCN)}.\hskip 1em plus 0.5em minus 0.4em\relax IEEE, 2018, pp. 1--9.

\bibitem{zhang2020radio}
S.~Zhang and R.~Zhang, ``Radio map-based {3D} path planning for cellular-connected {UAV},'' \emph{IEEE Transactions on Wireless Communications}, vol.~20, no.~3, pp. 1975--1989, 2020.

\bibitem{dong2022radio}
Y.~Dong, C.~He, Z.~Wang, and L.~Zhang, ``Radio map assisted path planning for {UAV} anti-jamming communications,'' \emph{IEEE Signal Processing Letters}, vol.~29, pp. 607--611, 2022.

\bibitem{chen2019efficient}
J.~Chen and D.~Gesbert, ``Efficient local map search algorithms for the placement of flying relays,'' \emph{IEEE Transactions on Wireless Communications}, vol.~19, no.~2, pp. 1305--1319, 2019.

\bibitem{zheng2022geography}
Y.~Zheng and J.~Chen, ``Geography-aware optimal {UAV} {3D} placement for {LOS} relaying: A geometry approach,'' 2022, available online: https://arxiv.org/abs/2209.15161.

\bibitem{zheng2023online}
{ Y. Zheng, Y. Wang, and J. Chen}, ``Online search for {UAV} relay placement for free-space optical communication under shadowing,'' \emph{Intelligent and Converged Networks}, vol.~4, no.~1, pp. 28--40, 2023\color{black}.

\bibitem{gesbert2022uav}
D.~Gesbert, O.~Esrafilian, J.~Chen, R.~Gangula, and U.~Mitra, ``{UAV}-aided {RF} mapping for sensing and connectivity in wireless networks,'' \emph{Wireless Communications}, pp. 1--7, 2022.

\bibitem{li2018uav}
{B. Li, Z. Fei, and Y. Zhang}, ``{UAV} communications for {5G} and beyond: Recent advances and future trends,'' \emph{IEEE Internet of Things Journal}, vol.~6, no.~2, pp. 2241--2263, 2018.

\bibitem{zeng2019accessing}
{Y. Zeng, Q. Wu, and R. Zhang}, ``Accessing from the sky: A tutorial on {UAV} communications for {5G} and beyond,'' \emph{Proceedings of the IEEE}, vol. 107, no.~12, pp. 2327--2375, 2019.

\bibitem{geraci2022will}
{ G. Geraci, A. Garcia-Rodriguez, M. M. Azari, A. Lozano, M. Mez- zavilla, S. Chatzinotas, Y. Chen, S. Rangan, and M. Di Renzo}, ``What will the future of {UAV} cellular communications be? a flight from {5G} to {6G},'' \emph{IEEE communications surveys \& tutorials}, vol.~24, no.~3, pp. 1304--1335, 2022\color{black}.

\bibitem{hashesh2022ai}
{ A. O. Hashesh, S. Hashima, R. M. Zaki, M. M. Fouda, K. Hatano, and A. S. T. Eldien,}, ``{AI}-enabled {UAV} communications: Challenges and future directions,'' \emph{IEEE Access}, vol.~10, pp. 92\,048--92\,066, 2022\color{black}.

\bibitem{al2014modeling}
A.~Al-Hourani, S.~Kandeepan, and A.~Jamalipour, ``Modeling air-to-ground path loss for low altitude platforms in urban environments,'' in \emph{IEEE Global Communications Conference (GLOBECOM)}, 2014, pp. 2898--2904.

\bibitem{wang2022ultra}
R.~Wang, M.~A. Kishk, and M.-S. Alouini, ``Ultra-dense {LEO} satellite-based communication systems: {A} novel modeling technique,'' \emph{Communications Magazine}, vol.~60, no.~4, pp. 25--31, 2022.

\bibitem{wang2022stochastic}
------, ``Stochastic geometry-based low latency routing in massive {LEO} satellite networks,'' \emph{IEEE Transactions on Aerospace and Electronic Systems}, vol.~58, no.~5, pp. 3881--3894, 2022.

\bibitem{esrafilian20173d}
O.~Esrafilian and D.~Gesbert, ``{3D} city map reconstruction from {UAV}-based radio measurements,'' in \emph{GLOBECOM 2017-2017 IEEE Global Communications Conference}.\hskip 1em plus 0.5em minus 0.4em\relax IEEE, 2017, pp. 1--6.

\bibitem{alzenad2019coverage}
M.~Alzenad and H.~Yanikomeroglu, ``Coverage and rate analysis for vertical heterogeneous networks ({VHetNets}),'' \emph{IEEE Transactions on Wireless Communications}, vol.~18, no.~12, pp. 5643--5657, 2019.

\bibitem{wackerly2014mathematical}
D.~Wackerly, W.~Mendenhall, and R.~L. Scheaffer, \emph{Mathematical statistics with applications}.\hskip 1em plus 0.5em minus 0.4em\relax Cengage Learning, 2014.

\bibitem{wu2020learning}
C.-P. Wu, Y.-R. Li, J.-L. Wang, H.-P. Lin, L.-C. Wang, S.-S. Jeng, and J.-Y. Chen, ``Learning-based downlink user selection algorithm for {UAV-BS} communication network,'' in \emph{2020 IEEE 17th Annual Consumer Communications \& Networking Conference (CCNC)}.\hskip 1em plus 0.5em minus 0.4em\relax IEEE, 2020, pp. 1--5.

\bibitem{ITU-R}
ITU-R, ``Propagation data and prediction methods required for the design of terrestrial broadband millimetric radio access systems operating in a frequency range of about $20-50$ {GHz},'' \emph{Radiowave propagation}, 2003 \color{black}.

\bibitem{lou2023coverage}
Z.~Lou, B.~E.~Y. Belmekki, and M.-S. Alouini, ``Coverage analysis of hybrid {RF/THz} networks with best relay selection,'' \emph{IEEE Communications Letters}, vol.~27, no.~6, pp. 1634--1638, 2023.

\bibitem{lou2024terrain}
Z.~Lou, R.~Wang, B.~E.~Y. Belmekki, M.~A. Kishk, and M.-S. Alouini, ``Terrain-based {UAV} deployment: Providing coverage for outdoor users,'' \emph{IEEE Transactions on Vehicular Technology}, 2024.

\bibitem{kishk2020aerial}
M.~Kishk, A.~Bader, and M.-S. Alouini, ``Aerial base station deployment in {6G} cellular networks using tethered drones: The mobility and endurance tradeoff,'' \emph{IEEE Vehicular Technology Magazine}, vol.~15, no.~4, pp. 103--111, 2020.

\bibitem{belmekki2022unleashing}
B.~E.~Y. Belmekki and M.-S. Alouini, ``Unleashing the potential of networked tethered flying platforms: {P}rospects, challenges, and applications,'' \emph{IEEE Open Journal of Vehicular Technology}, vol.~3, pp. 278--320, 2022.

\bibitem{lou2023haps}
Z.~Lou, B.~E.~Y. Belmekki, and M.-S. Alouini, ``{HAPS} in the non-terrestrial network nexus: Prospective architectures and performance insights,'' \emph{IEEE Wireless Communications}, vol.~30, no.~6, pp. 52--58, 2023.

\bibitem{lou2021green}
Z.~Lou, A.~Elzanaty, and M.-S. Alouini, ``Green tethered {UAV}s for {EMF}-aware cellular networks,'' \emph{IEEE Transactions on Green Communications and Networking}, vol.~5, no.~4, pp. 1697--1711, 2021.

\end{thebibliography}

\end{document}